\pdfoutput=1
\documentclass[aps,prl,twocolumn,superscriptaddress,nofootinbib]{revtex4}

\usepackage{amssymb,amsfonts,amsmath,amsthm}
\usepackage{color,enumerate,graphicx,hyperref}

\newcommand{\eps}{\varepsilon}
\newcommand{\R}{\mathbb R}

\newcommand{\m}{\mathbf{m}}

\newcommand{\rr}{\mathbf{r}}

\newcommand{\pp}{\mathbf{p}}

\newcommand{\mpa}{m_\parallel}

\newcommand{\mpe}{\mathbf{m}_\perp}

\begin{document}

\title{\bf A micromagnetic theory of skyrmion lifetime in ultrathin
  ferromagnetic films}

\author{ Anne Bernand-Mantel}

\affiliation{Universit\'e de Toulouse, Laboratoire de Physique et
  Chimie des Nano-Objets, UMR 5215 INSA, CNRS, UPS, 135 Avenue de
  Rangueil, F-31077 Toulouse Cedex 4, France}

\email{bernandm@insa-toulouse.fr}

\author{Cyrill B. Muratov}

\affiliation{Department of Mathematical Sciences, New Jersey Institute
  of Technology, Newark, New Jersey 07102, USA}

\email{muratov@njit.edu}


\author{Valeriy V. Slastikov}

\affiliation{School of Mathematics, University of Bristol, Bristol BS8
  1TW, United Kingdom}

\email{Valeriy.Slastikov@bristol.ac.uk}

\date{\today}

\begin{abstract}
  We use the continuum micromagnetic framework to derive the formulas
  for compact skyrmion lifetime due to thermal noise in ultrathin
  ferromagnetic films with relatively weak interfacial
  Dzyaloshinskii-Moriya interaction.  In the absence of a saddle point
  connecting the skyrmion solution to the ferromagnetic state, we
  interpret the skyrmion collapse event as ``capture by an absorber''
  at microscale. This yields an explicit Arrhenius collapse rate with
  both the barrier height and the prefactor as functions of all the
  material parameters, as well as the dynamical paths to collapse.
\end{abstract}

\maketitle

\paragraph{Introduction.}

Magnetic skyrmions are a characteristic example of topological
solitons existing at nanoscale. Their extensive studies in the past
ten years revealed a very rich underlying physics as well as potential
applications in the field of spintronics
\cite{kiselev11,nagaosa13,fert17,zhang20}.  While the fundamental
object for applications is an individual skyrmion in a homogeneous
ferromagnetic environment, for topological reasons it cannot be
created or annihilated by a continuous transformation from the
ferromagnetic state.  This transition is however enabled by the
discrete nature of the condensed matter as observed experimentally
\cite{sampaio13,romming13,hagemeister15,wild17}.

The detailed physical mechanisms of skyrmion annihilation have been
investigated at the nanoscale using atomic spin simulations combined
with methods of finding the minimum energy path and harmonic
transition state theory
\cite{bessarab15,lobanov16,cortes-ortuno17,bessarab18,desplat18,
  heil19, lobanov21}.
In particular, the energy barrier $\Delta E$ separating the skyrmion
state from the ferromagnetic state was obtained numerically for some
given sets of parameters and the skyrmion annihilation rate was
estimated by a simple Arrhenius law $\Gamma_0 e^{-\Delta E/(k_B T)}$,
where $\Gamma_0$ is the rate prefactor, also called the attempt
frequency.  While early works used standard values of $\Gamma_0$ which
are in the range from $10^9$ to $10^{12}$ Hz in the macrospin model
\cite{brown63}, more recent studies show that in the case of skyrmions
$\Gamma_0$ can vary by many orders of magnitude
\cite{wild17, vonmalottki19, desplat20,hoffmann20}.

Despite this progress, there are limitations to the atomistic
simulations. First, they are computationally expensive, which limits
the accessible skyrmion sizes (usually below 5 nm in diameter) and the
physical parameter ranges that can be explored. Second, the obtained
results depend on the microscopic details that are not necessarily
known or controlled in the case of nanocrystalline systems. Under
these circumstances, there is clearly a need for a more coarse-grained
theory that would provide universal relations between the skyrmion
lifetime and the material parameters. Moreover, it is reasonable to
expect that under many physically relevant conditions the microscopic
details do not play a dominant role for fluctuation-driven skyrmion
collapse. For example, the skyrmion size is often much larger than the
atomic lattice spacing when it loses its topological protection via
the disappearance of its core \cite{verga14,heil19}.

In this Letter, we develop a theory of skyrmion lifetime based on the
continuum field theory and derive the expressions for both the energy
barrier and the attempt frequency as functions of all the material
parameters. Starting with the stochastic Landau-Lifshitz-Gilbert
partial differential equation, we first derive several integral
identities associated with the fundamental continuous symmetry groups
of the exchange energy. Then, in the exchange dominated regime, we
carry out a finite-dimensional reduction of the stochastic skyrmion
dynamics and obtain a system of stochastic ordinary differential
equations for the skyrmion radius and angle. Finally, in the small
thermal noise regime we use the obtained equations to calculate the
Arrhenius rate, including the prefactor, by interpreting the skyrmion
collapse event as ``capture by an absorber'' for the skyrmion radius
at the atomic scale.

\paragraph{Model.} At the continuum level the magnetization dynamics
in an ultrathin ferromagnetic film at finite temperature is described
by the stochastic Landau-Lifshitz-Gilbert (sLLG) equation
\cite{landau8,garcia07,brown63,garcia-palacios98} (for the technical
details on all the formulas in the paper, see \cite{suppl})
\begin{align}
  \label{LLG} 
\frac{\partial \mathbf{m}}{\partial t}
  =  -\m\times\mathbf{h}_{\rm{eff}} + {\alpha} \mathbf{m}\times
    \frac{\partial \mathbf{m}}{\partial t},
\end{align}
where $\m = \m(\rr, t)$ is the unit magnetization vector at position
$\rr \in \R^2$ measured in the units of the exchange length
$\ell_{ex} = \sqrt{2 A / (\mu_0 M_s^2)}$, where $A$ is the exchange
stiffness, $M_s$ is the saturation magnetization and $\mu_0$ is vacuum
permeability, and time $t$ measured in the units of
$\tau_0 = (\gamma \mu_0 M_s)^{-1}$, where $\gamma$ is the gyromagnetic
ratio, $\alpha$ is the dimensionless Gilbert damping parameter, and
$\mathbf{h}_{\rm{eff}} = \mathbf{h}_{\rm{eff}}(\rr, t) \in \R^3$ is
the effective field given by
\begin{align}
  \label{heff}
  \mathbf{h}_{\rm{eff}} = - \frac{\delta E(\m)}{\delta \m} + \sqrt{2
  \alpha \eps} \, \boldsymbol{\xi},
\end{align}
where $E(\m)$ is the micromagnetic energy measured in the units of
$2Ad$, with $d$ being the film thickness, and
$\boldsymbol{\xi} = \boldsymbol{\xi}(\mathbf r, t) \in \R^3$ is a
suitable regularization of a three-dimensional delta-correlated
spatiotemporal white noise \cite{dapratozybczik}.  In the local
approximation for the stray field and in the absence of the applied
field we have \cite{
  bogdanov89,bogdanov94,thiaville12,ms:prsla17,bms:prb20, bms:arma21}
\begin{align}
  E(\m) \!
  = \! \frac12 \! \int_{\R^2} \! \Big\{ |\nabla \m|^2 \! + \! (Q
  - 1) |\mpe|^2 \! - \! 2 \kappa
  \mpe \! \cdot \! \nabla \mpa  \! \Big\} d^2 r,
    \label{Eh}
\end{align}
which consists of, in order of appearance, the exchange, the effective
uniaxial out-of-plane anisotropy ($Q > 1$) and the interfacial
Dzyaloshinskii-Moriya interaction (DMI) terms, respectively. Above we
defined $\mpe \in \R^2$ and $\mpa \in \R$ to be the respective
in-plane and out-of-plane components of the magnetization vector
$\m = (\mpe, \mpa)$, and introduced
\begin{align}
  \label{Qkappa}
  Q = {2 K \over \mu_0 M_s^2}, \qquad \kappa = D \sqrt{2 \over \mu_0
  M_s^2 A}, \qquad   \eps = {k_B T \over 2 A d},
\end{align}
where $K$ is the magnetocrystalline uniaxial anisotropy constant, $D$
is the DMI constant and $k_B T$ is temperature in the energy
units. The dimensionless parameters in \eqref{Qkappa} characterize the
anisotropy, the DMI and the noise strengths, respectively.

\paragraph{Integral identities.} We begin by rewriting the sLLG
equation in the spherical coordinates, setting
$\m =(\sin\theta \cos\phi, \sin\theta \sin\phi, \cos\theta)$, and
express it in terms of $\theta$ and $\phi$. After some tedious
algebra, we get
\begin{widetext}
  \begin{align}
  \label{thphi}
  \left(
    \begin{array}{cc}
      \alpha & -1\\
      1 & \alpha
    \end{array} \right) 
  \left(
    \begin{array}{c}
      \theta_t \\
      \sin\theta \phi_t
    \end{array} \right)
  = 
  \left(
    \begin{array}{c}
      \Delta \theta - \sin\theta \cos\theta |\nabla \phi|^2 -
      (Q-1) \cos\theta \sin\theta +\kappa \, \sin^2\theta
      \nabla\phi 
      \cdot \pp + \sqrt{2 \alpha \eps} \, \eta \\
      \sin\theta \Delta
      \phi  + 2 \cos\theta \nabla\theta \cdot \nabla\phi - \kappa
      \sin\theta \nabla\theta\cdot \pp + \sqrt{2 \alpha \eps} \, \zeta
    \end{array}
  \right), 
\end{align}
\end{widetext}
where $\eta$ and $\zeta$ are two independent, delta-correlated
spatiotemporal white noises, $\pp = (-\sin \phi, \cos \phi)$, and here
and everywhere below the letter subscripts denote partial derivatives
in the respective variables.

We next derive several integral identities for the solutions of
\eqref{thphi} that will be useful in obtaining the evolution equations
for the skyrmion characteristics. These identities are closely related
to the continuous symmetry groups associated with the exchange energy
term, which dominates in the considered regime.  We start with the
group of rotations and scalar multiply \eqref{thphi} by
$(0, \sin \theta)$. A subsequent integration over space yields
\begin{multline}
  \! \! \int_{\R^2}\sin\theta \theta_t \, d^2 r \! + \! \alpha
    \int_{\R^2}\sin^2\theta \phi_t \, d^2 r  \! + \! \kappa
    \int_{\R^2} 
    \sin^2\theta \nabla\theta \! \cdot \! \pp \, d^2 r  \\ 
  = \sqrt{2 \alpha \eps}
    \left( \int_{\R^2} \sin^2 \theta\, d^2 r\right)^{1/2}
    \dot{W}_1(t),  \label{W1}
\end{multline}
where $W_1(t)$ is a Wiener process, and the dot denotes the time
derivative. Here we noted that an integral of a
divergence term vanishes for the profiles that approach a constant
vector at infinity.

Now we use the group of dilations and scalar multiply \eqref{thphi} by
$(\nabla \theta \cdot (\mathbf r - \mathbf r_0(t)), \sin \theta \nabla
\phi \cdot (\mathbf r - \mathbf r_0(t)))$, where $\mathbf r_0(t)$ is
arbitrary. This yields
\begin{multline}
  \! \! - \! \int_{\R^2} \!  (\rr \! -\! \rr_0(t)) \! \cdot \! \nabla
  \theta\, \sin\theta \phi_t d^2 r \! + \! \alpha \int_{\R^2} \!  (\rr
  \!  - \!  \rr_0(t) \!) \! \cdot \!  \nabla \theta\, \theta_t d^2 r \\
  \! + \!  \alpha \int_{\R^2} \! \sin^2\theta \nabla\phi \cdot (\rr \!
  - \!  \rr_0(t) \! ) \phi_t d^2 r \! + \! \int_{\R^2} \! (\rr \! - \!
  \rr_0(t) \! ) \cdot \! \nabla \phi\, \theta_t \sin\theta d^2 r \\ \!
  = \!  \int_{\R^2} \!  (\rr \! - \! \rr_0(t) \! ) \! \cdot \! \nabla
  \theta\, (-(Q \! -\!  1) \cos\theta \sin\theta \! +\! \kappa \,
  \sin^2\theta \nabla\phi \! \cdot \! \pp) d^2 r \\ \!  + \!
  \int_{\R^2} \! \sin\theta \nabla\phi \! \cdot \!  (\rr \! -\!
  \rr_0(t) \!)( -\kappa \sin\theta \nabla\theta \! \cdot \! \pp ) d^2 r
  \\ \! + \! \sqrt{\int_{\R^2} \! | (\rr \! - \!  \rr_0(t) \!) \!  \cdot
    \!  \nabla \theta|^2 d^2 r \! + \! \int_{\R^2} \! \sin^2 \theta
    |(\rr \! - \! \rr_0(t) \!) \! \cdot \! \nabla \phi|^2 d^2 r} \\
  \times \sqrt{2 \alpha \eps} \, \dot{W}_2(t) , \label{W2}
\end{multline}
where $W_2(t)$ is another Wiener process.  Finally, we use the
translational symmetries of the exchange energy and scalar multiply
the stochastic LLG equation by $(\theta_x, \sin \theta \phi_x)$ or
$(\theta_y, \sin \theta \phi_y)$ to obtain two similar identities
involving two other Wiener processes $W_3(t)$ and $W_4(t)$
\cite{suppl}.  Note that in general the Wiener processes $W_1(t)$
through $W_4(t)$ are not independent.

\paragraph{Reduction to a finite-dimensional system.} To proceed
further, we focus on the regime in which a good approximation to the
solutions of the sLLG equation may be obtained by means of a matched
asymptotic expansion. This regime, in which
$0 < \kappa \ll \sqrt{Q - 1}$ gives rise to a skyrmion profile
$(\theta, \phi) = (\bar\theta, \bar \phi)$ whose radius $\rho_0$ is
asymptotically \cite{bms:prb20,komineas20,bms:arma21}
\begin{align}
  \label{rho0}
  \rho_0 \simeq {\kappa \over 2 (Q - 1) \ln \left( a \kappa^{-1}
  \sqrt{Q - 1} \right)},
\end{align}
for some $a \sim 1$. It is characterized by a compact core on the
scale of $\rho_0$:
\begin{align}
  \label{in}
  \bar \theta(r) \simeq 2 \arctan (r / \rho_0),
\end{align}
which is the Belavin-Polyakov profile \cite{belavin75} that minimizes
the exchange energy at leading order, and an exponentially decaying
tail on the scale of the Bloch wall length $L = (Q - 1)^{-1/2}$:
\begin{align}
  \label{out}
  \bar \theta(r) \simeq \pi - 2 \rho_0 \sqrt{Q - 1} \,  K_1(r \sqrt{Q
  - 1}), 
\end{align}
where $K_0(z)$ is the modified Bessel function of the second kind,
that minimizes the exchange plus anisotropy energy to the leading
order. In both the core and the tail $\bar \phi = \psi - \pi$, where
$x = r \cos \psi$ and $y = r \sin \psi$ are the polar coordinates
relative to the skyrmion center.

Dynamically, one would expect that for $\alpha \sim 1$ the above
profile would stabilize on the diffusive timescale
$\tau_\mathrm{core} \sim \rho^2$ in the core, and on the relaxation
timescale $\tau_\mathrm{relax} \sim (Q - 1)^{-1}$ in the tail,
respectively. Therefore, on the timescale
$\tau_\mathrm{relax} \gtrsim \tau_{core}$ the dynamical profile
$\theta(\rr, t)$ in the skyrmion core would be expected to be
dominated by the exchange and, therefore, stay close to a suitably
translated, rotated and dilated Belavin-Polyakov profile:
\begin{align}
  \theta(\rr, t) & \simeq 2 \arctan (|\rr - \rr_0(t)| / \rho(t)), \\
  \phi(\rr, t) & \simeq \arg (\rr - \rr_0(t)) - \pi + \varphi(t).
\end{align}
Similarly, on the timescale $\gtrsim \tau_\mathrm{relax}$ the skyrmion
profile in the tail should approach
\begin{align}
  \theta(\rr, t) \simeq \pi - 2 \rho(t) \sqrt{Q - 1} \,  K_1(|\rr -
  \rr_0(t)| \sqrt{Q - 1}).
\end{align}
Here, the functions $\rho(t)$, $\varphi(t)$ and $\rr_0(t)$ may be
interpreted, respectively, as the instantaneous radius, rotation
angle and the center of the skyrmion.

The above approximate solution may be substituted into our integral
identities to obtain a closed set of equations for $\rho(t)$,
$\varphi(t)$ and $\rr_0(t) = (x_0(t), y_0(t))$:
\begin{multline}
  \label{rhophi}
 \!\!\!  {d \over dt} \left(
  \begin{array}{c}
    \ln \rho \\
   \varphi
  \end{array}
  \right) = - {1 \over 1 + \alpha^2}
  \left(
  \begin{array}{cc}
    \alpha & -1 \\
    1 & \alpha
  \end{array}
\right)  \left(
        \begin{array}{c}
          Q - 1 -
          {\kappa \cos \varphi \over 2 \rho \ln (L / \rho)} \\
          {\kappa \sin\varphi  \over 2 \rho \ln (L/\rho)}
        \end{array}
      \right) \\
      + \sqrt{\alpha \eps \over 4 \pi (1 + \alpha^2) \rho^2 \ln (L /
        \rho)} \left(
  \begin{array}{c}
    \dot W_1(t) \\
    \dot W_2(t) 
  \end{array}
  \right),
\end{multline}
and
\begin{align}
  \dot x_0
  = \sqrt{\alpha \eps \over 2 \pi (1 + \alpha^2)} \, \dot 
  W_3, \quad
  \dot y_0
  = 
  \sqrt{\alpha \eps \over 2 \pi (1 + \alpha^2)} \, \dot W_4.  
\end{align}
Furthermore, to the leading order the Wiener processes $W_1(t)$
through $W_4(t)$ are all mutually independent. It is understood that
$\rho$ is bounded above by some $L_0 < L$. Moreover, when
$\rho \sim \rho_0$, we may set the large logarithmic factor
$\ln (L / \rho)$ to a constant
$\Lambda = \ln (a \kappa^{-1} \sqrt{Q - 1})$ to the leading order.
Introducing the new variable
$\bar z = \bar x + i \bar y = \rho e^{i \varphi}$ then results in the
following stochastic differential equation:
\begin{align}
  \label{dz}
  d\bar z(t) = -{\alpha + i \over 1 + \alpha^2} \left[ (Q - 1)
  \bar z(t) - {\kappa \over 2 \Lambda} \right] dt \notag \\
  + \sqrt{\alpha \eps
  \over 4 \pi \Lambda (1 + \alpha^2)} \, d\bar W(t),
\end{align}
where $\bar W(t) = W_1(t) + i W_2(t)$ is a complex-valued Wiener
process. Note that the dynamics of $\bar z(t)$ decouples from that of
$\mathbf r_0(t)$, with the latter undergoing a simple diffusion with
diffusivity
$D_\mathrm{eff} = {\alpha \eps \over 4 \pi (1 + \alpha^2)}$, in
agreement with \cite{schutte14}.

\paragraph{Calculation of the collapse rate.} We now focus on the
analysis of \eqref{dz}. It describes a two-dimensional shifted
Ornstein-Uhlenbeck process, whose equilibrium measure is given by the
Boltzmann distribution
\begin{align}
  p_\mathrm{eq}(\bar z) = {4 \Lambda \eps^{-1} (Q - 1)} e^{-{H(\bar z)
  \over \eps}},  
\end{align}
where
\begin{align}
  H(\bar z) = 4 \pi \Lambda (Q - 1) |\bar z -
  \bar z_0|^2, \quad \bar z_0 = {\kappa \over 2 \Lambda (Q - 1)},
\end{align}
which is peaked around $\bar z = \bar z_0$ in the complex plane.  This
distribution is attained on the timescale of $\tau_\mathrm{relax}$.

\begin{figure*}
  \centering
  \includegraphics[width=7in]{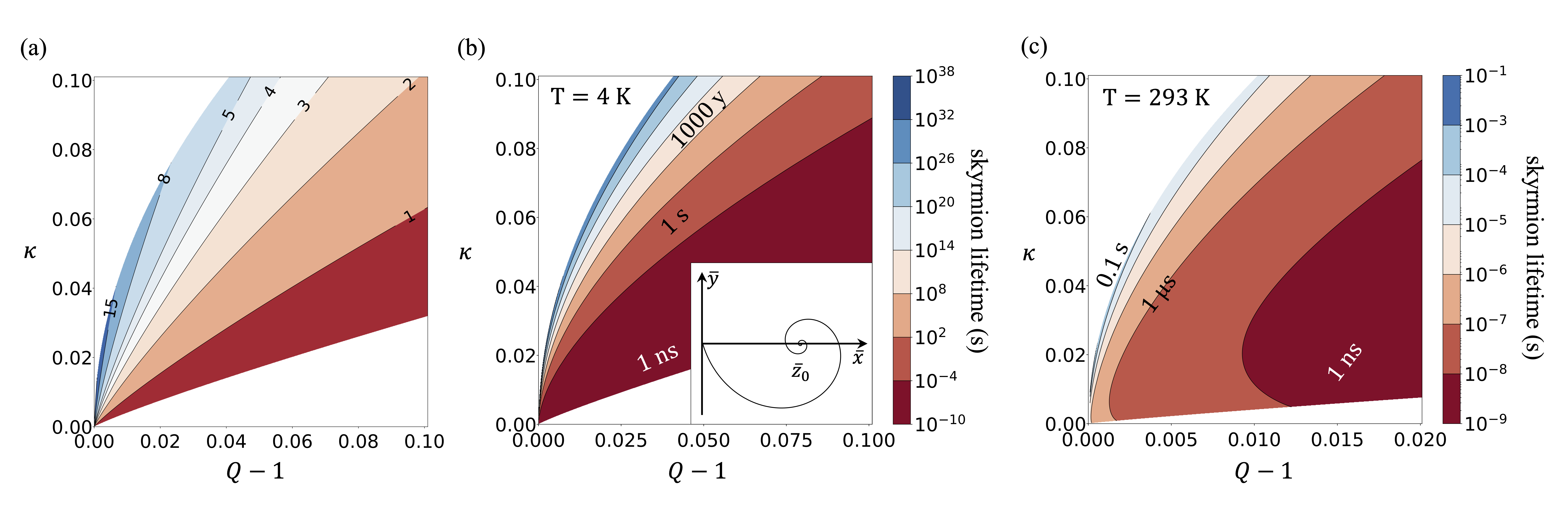}
  \caption{(a) Skyrmion equilibrium radius $\ell_\mathrm{ex} \rho_0 $
    in nm, with $\ell_\mathrm{ex} = 8.4$ nm and $\rho_0$ from
    \eqref{rho0}. (b) and (c) Skyrmion lifetime $\tau_0 / J_\delta$
    for $\tau_0 = 7.8 \times 10^{-12}$ s, $\alpha = 0.3$,
    $\delta = 0.0475$, with $J_\delta$ from \eqref{J0} for
    $\eps = 0.0046$ ($T= 4$K) in (b) and $J_\delta$ from \eqref{J10}
    for $\eps = 0.337$ ($T= 293$K) in (c). In (a) and (b), only the
    region $\kappa / \sqrt{Q - 1} < \frac12$ is shown.  Inset: The
    optimal collapse path $\bar z_\mathrm{opt}(t)$ from \eqref{zopt}.
  }
  \label{fig:fig}
\end{figure*}

Notice that the probability of the solutions of \eqref{dz} starting at
$\bar z = \bar z_0$ to hit the origin is zero, although the
probability to come to an arbitrarily small neighborhood of the origin
is unity. Therefore, within \eqref{dz} a more careful definition of
the skyrmion collapse event is necessary. For that purpose, we note
that when the skyrmion radius becomes sufficiently small, the
continuum micromagnetic description of the magnetization profile
breaks down. This happens when the skyrmion radius reaches the atomic
scale, at which point the skyrmion loses its topological protection.
Therefore, to model skyrmion collapse we supplement \eqref{dz} with an
absorbing boundary condition at $|\bar z| = \delta$ for some cutoff
radius $\delta \ll \bar z_0$. In atomically thin films, this cutoff
radius is on the order of the film thickness measured in the units of
the exchange length, $\delta \sim d / \ell_\mathrm{ex}$. The mean
skyrmion lifetime may then be found by solving an appropriate boundary
value problem in the plane. When $\eps \ll 1$, it may be obtained by
investigating the stationary solution $p = p(\bar x, \bar y)$ of the
Fokker-Planck equation associated with \eqref{dz} with a suitable
source $g(\bar x, \bar y)$ away from the absorber:
\begin{align}
  \left[ (Q - 1) x - {\kappa \over 2
  \Lambda} \right]  (\alpha q_{\bar x}  - q_{\bar y}) + y (Q - 1)
  (\alpha q_{\bar y} + q_{\bar x} ) \notag \\
  = {\alpha \eps \over 8 \pi
  \Lambda} ( q_{\bar x \bar x} + q_{\bar y 
  \bar y} ) + g p_\mathrm{eq}^{-1}, \label{q}
\end{align}
where we introduced $q = p / p_\mathrm{eq}$ and set $q = 0$ at
$|\bar z| = \delta$. When $\eps \to 0$, we are interested in the
boundary layer solution of \eqref{q} in the neighborhood of
$(\bar x, \bar y) = (\delta, 0)$, in which the probability flux is
concentrated \cite{gardiner}. The function $q$ approaches 1 away from
the boundary layer. Then to the leading order in $\eps$, the solution
in a small neighborhood around the point
$(\bar x, \bar y) = (\delta, 0)$ is given by
\begin{align}
  q(\bar x, \bar y) \simeq 1 - \exp \left\{ -{8 \pi \Lambda \over
  \eps} \left[ (Q - 1) \delta -  {\kappa \over 2 \Lambda} \right]
  (\bar x - \delta) \right\}. 
\end{align}
Assuming that $\eps \ll \kappa \delta$, the corresponding total
probability flux into the boundary is, to the leading order,
\begin{align}
  J_\delta = {\alpha \eps \over 8 \pi \Lambda(1 + \alpha^2)}
  \int_{|\bar z| = 
  \delta} p_\mathrm{eq} |\nabla q| \, ds  \\
  \simeq  {\alpha \Lambda (Q - 1) \over
  1 + \alpha^2} \left[ {\kappa \over 2 \Lambda} - (Q - 1) \delta
  \right] \left( {8 \delta \over \eps \kappa} \right)^{1/2} \,
  e^{-{H(\delta) 
  \over \eps}}, 
\end{align}
which has the form of an Arrhenius law with explicit expressions for
the barrier height $H(\delta)$ and the prefactor. Notice that the
latter depends weakly on the parameter $\delta \ll 1$, and to the
leading order we have
\begin{align}
  \label{J0}
  J_\delta \simeq {\alpha (Q - 1) A_\delta \over 1 + \alpha^2} \left( {
  2 \kappa \delta
  \over \eps} \right)^{1/2}  \exp \left\{ {-{\pi \kappa^2 \over
  \eps \Lambda (Q - 1)}} \right\},
\end{align}
where $\pi \kappa^2 / [\Lambda (Q - 1)]$ is the leading order barrier
height $H(0)$ and
\begin{align}
  \label{Adelta}
  A_\delta = \exp \left\{ {4 \pi \kappa \delta \over \eps} \left[ 1 -
  {\Lambda (Q - 1) \delta \over \kappa} \right] \right\} > 1
\end{align}
is an anomalous factor due to a small reduction
$\Delta H = H(0) - H(\delta)$ of the barrier height resulting from the
presence of the absorber at microscale needed to break the topological
protection.

The quantity in \eqref{J0} gives the leading order asymptotic skyrmion
collapse rate for $\delta \ll 1$ as $\eps \to 0$.  The exponential
term is nothing but the Arrhenius factor associated with the energy
barrier to collapse, to the leading order in
$\kappa / \sqrt{Q - 1} \ll 1$. A comparison with the result of the
numerical solution for the radial skyrmion profile \cite{suppl} shows
that taking $a = 2.8$ in the definition of
$\Lambda = \ln (a \kappa^{-1} \sqrt{Q - 1})$ reproduces the exact
barrier height to within 17\% for all $\kappa / \sqrt{Q - 1} < 0.8 $.

It is also possible to obtain the skyrmion collapse rate in the limit
$\delta \to 0$ with $\eps \ll 1$ and all the other parameters fixed,
corresponding to the opposite extreme $\eps \gg \kappa \delta$. Here
in the $O(\eps/\kappa)$ neighborhood of the absorber the function
$q(\bar x, \bar y)$ is, to the leading order in $\delta \ll 1$,
\begin{align}
  q(\bar x, \bar y) \simeq {\ln \left( {\sqrt{\bar x^2 + \bar y^2}
  \over \delta} \right) \over \ln \left( {b \alpha \eps \over \kappa  
  \delta \sqrt{1 + \alpha^2}}
  \right)},
\end{align}
where $b \approx 0.179$.  An analogous computation to the one leading
to \eqref{J0} yields in this case
\begin{align}
  \label{J10}
  J_\delta \simeq {\alpha (Q-1) A_\delta \over (1 + \alpha^2) \ln
  \left( { b \alpha \eps \over \kappa  
  \delta \sqrt{1 + \alpha^2} } \right)} \,  \exp
  \left\{ {-{\pi \kappa^2 \over \eps \Lambda (Q - 
  1)}} \right\}.    
\end{align}
The condition $\eps \gg \kappa \delta$ or, equivalently,
$\delta \ll \eps / \kappa$ ensures that
$p_\mathrm{eq}(\bar x, \bar y)$ does not vary appreciably across the
absorber boundary, making $A_\delta \simeq 1$ as $\delta \to 0$.

\paragraph{Skyrmion collapse paths.} The dynamics of skyrmion collapse
in the small noise limit may be understood through the minimization of
the large deviation action associated with \eqref{dz}
\cite{freidlin}:
\begin{multline}
  S = {2 \pi \Lambda (1 + \alpha^2) \over \alpha} \\
  \times \int_0^T \left|\dot{\bar z} + {\alpha + i \over
      1 + \alpha^2} \left[ (Q - 1) \bar z - {\kappa \over 2 \Lambda}
    \right] \right|^2 dt. \label{action}
\end{multline}
Minimizing over all trajectories $\bar z(t)$ that start at
$\bar z(0) = \bar z_0$ and terminate at $\bar z(T) = \delta$, and then
sending $T \to \infty$, one obtains the optimal collapse trajectory
$\bar z = \bar z_\mathrm{opt}(t)$, where to the leading order in
$\delta \ll 1$ we have
\begin{align}
  \label{zopt}
  \bar z_\mathrm{opt}(t) = \bar z_0 \left( 1 - e^{{\alpha - i \over 1
  + \alpha^2} (Q - 1) (t - T)} \right). 
\end{align}
As expected, for $\alpha \lesssim 1$ the collapse occurs on the
timescale $\alpha^{-1} (Q - 1)^{-1}$ and acquires an oscillatory
character for $\alpha \ll 1$. The optimal path to collapse is
illustrated in the inset in Fig. \ref{fig:fig}(b). Notice that for
$\alpha \ll 1$ the skyrmion angle rotates as the skyrmion radius
shrinks to zero, similarly to what is observed in current-driven
skyrmion collapse \cite{verga14}.

\paragraph{Parametric dependence of skyrmion lifetime.} We now use the
obtained formulas for the collapse rate to calculate the skyrmion
lifetimes as functions of the material parameters in a typical
ultrathin ferromagnetic film. For that purpose, we take the same
parameters as in Sampaio \emph{et. al.} \cite{sampaio13}: $d = 0.4$
nm, $A = 15$ pJ/m, $M_s = 0.58$ MA/m, $\alpha = 0.3$. This yields
$\ell_\mathrm{ex} = 8.4$ nm, $\tau_0 = 7.8 \times 10^{-12}$ s and
$\delta = d/\ell_\mathrm{ex} = 0.0475$. The equilibrium skyrmion
radius and lifetime $\tau_0 / J_\delta$ as functions of the
dimensionless parameters $Q - 1$ and $\kappa$ are plotted in
Fig. \ref{fig:fig}. The low temperature regime corresponding to
\eqref{J0} is illustrated in Fig. \ref{fig:fig}(b), while the high
temperature regime corresponding to \eqref{J10} is illustrated in
Fig. \ref{fig:fig}(c).  In both cases, the lifetime varies by many
orders of magnitude, and this variation is dominated by the
exponential dependence on the barrier height proportional to
$\bar \kappa^2$, where $\bar \kappa=\kappa/\sqrt{Q-1}$ is the
classical parameter which determines the transition from the
ferromagnetic to the helical ground state happening at
$\bar \kappa = 4/\pi$ \cite{bogdanov94}. The lifetime increases upon
increase of $\bar \kappa$ and is maximal when $\bar \kappa \sim 1$, at
the borderline of applicability of our analysis.

In addition to the variation of the barrier height, we also predict a
variation of the effective rate prefactor. This variation is stronger
in the low temperature case [Fig. \ref{fig:fig}(b)], where it is
dominated by the anomalous factor $A_\delta $. In this regime the rate
prefactor becomes strongly dependent on the reduction in the barrier
height due to the microscopic processes associated with the loss of
the topological protection modeled by us by an absorbing boundary
condition. This strong variation of the prefactor is similar to the
strong prefactor dependence on microscopic details (layer stacking,
number of magnetic monolayers, etc.) observed in recent simulations
\cite{hoffmann20}. In the high temperature regime
($\eps \gg \kappa \delta$), our prediction confirms that the prefactor
becomes essentially independent of the microscopic details. The
remaining dependence of the prefactor is dominated by its dependence
on $Q-1$ due to the expected proportionality of the attempt frequency
to the precession frequency $\gamma \mu_0 M_s(Q-1)$ \cite{brown63}.


\paragraph{Conclusion.}

To summarize, we carried out a derivation of skyrmion lifetime, using
the stochastic Landau-Lifshitz-Gilbert equation within the framework
of continuum micromagnetics and accounting for the loss of topological
protection via an absorbing boundary condition at microscale. Our
formulas in \eqref{J0} and \eqref{J10} provide the first relation of
skyrmion collapse rate to material parameters and could be used as a
guide in material system design in view of optimizing the skyrmion
lifetime for applications. The methodology developed by us may also
have a wide applicability to other physical systems in which a
topological defect disappears through singularity formation at the
continuum level.

\begin{acknowledgements}
  A.B.-M. acknowledges support from the DARPA TEE program through
  Grant MIPR No. HR0011831554. The work of C.B.M. was supported, in
  part, by NSF via grant DMS-1908709.  V. V. S. acknowledges support
  by Leverhulme grant RPG-2018-438.
\end{acknowledgements}

\bibliography{../../mura,../../nonlin,../../stat}

\end{document}